\documentclass[onecolumn,showpacs,preprintnumbers,amsmath,amssymb]{revtex4}
\usepackage{graphicx}
\usepackage{dcolumn}
\usepackage{bm,latexsym}
\usepackage{mathrsfs}

\def\Journal#1#2#3#4{{#1} {\bf #2}, #3 (#4)}

\def\PLA{{Phys. Lett.}  A}
\def\PLB{{Phys. Lett.}  B}
\def\PRL{Phys. Rev. Lett.}

\def\PRD{{Phys. Rev.} D}
\def\PR{{Phys. Rep.}}

\def\CQG{{Class. Quant. Grav.}}
\def\JMP{{J. Math. Phys.}}

\def\IJMPD{{Int. Jour. Mod. Phys.}D}
\def\CMP{{Commun. Math. Phys.}}
\def\JPA{{J. Phys.}  A}
\def\JMP{{J. Math. Phys.}}
\def\be{\begin{equation}}
\def\ee{\end{equation}}
\def\bea{\begin{eqnarray}}
\def\eea{\end{eqnarray}}

\def\sch{Schwarzschild }
\def\ibid{{\it ibid.}}

\begin{document}
\preprint{APS/123-QED}

\title{A simple theorem to generate exact black hole solutions}
\author{ Marcelo Salgado} 
\email{marcelo@nuclecu.unam.mx}
\homepage{http://www.nuclecu.unam.mx/~marcelo}
\affiliation{Instituto de Ciencias Nucleares 
\\ Universidad Nacional Aut\'onoma de M\'exico 
\\ Apdo. Postal 70--543 M\'exico 04510 D.F., M\'exico}

\date{\today}

\begin{abstract}
Under certain conditions imposed on the energy-momentum tensor, a 
theorem that characterizes a two-parameter family of static and 
spherically symmetric solutions to Einstein's field equations (black holes), 
is proved. A discussion on the asymptotics, 
regularity, and the energy conditions is provided. Examples that include the 
best known exact solutions within these symmetries are considered. 
A trivial extension of the theorem includes the cosmological 
constant {\it ab-initio}, providing then a three-parameter family of solutions. 
\end{abstract}

\pacs{04.20.Jb,04.70.Bw,04.40.Nr}

\maketitle

\section{Introduction}
Static and spherically symmetric spacetimes are one of the simplest 
kind of spacetimes that one can imagine in general relativity. Yet, 
even in this simple situation, solving the Einstein field equations may 
be far from trivial. The difficulty relies mainly on the nature of the 
energy-momentum tensor considered. At this regard, only a few exact 
solutions are known. Most of them represent hairless black 
hole (BH) solutions, like the \sch$\!\!$, Reissner-Nordstrom (RN) and 
\sch$\!\!$-de Sitter/anti-de Sitter (SdS/SAdS) solutions, for the vacuum, electromagnetic and cosmological constant sources 
respectively. The first two, are asymptotically flat while the 
latter is not. In fact, the uniqueness theorems of BH (with asymptotically 
flat boundary conditions) in Einstein-Maxwell (EM)
theory establishes that the unique BH solutions are stationary and 
axially symmetric and 
contained within the three-parameter Kerr-Newman family 
\cite{uniqueness}. The parameters are 
identified with the {\it mass}, {\it electric charge} and {\it angular momentum} 
of the BH. Therefore, in the 
static and spherically symmetric limit, this corresponds to the \sch$\!\!$-RN 
two-parameter family of solutions. 

When the energy-momentum tensor includes more complicated fields, uniqueness theorems 
or even solutions are difficult to obtain. However, other theorems 
\cite{nohair,eloy} forbid the existence of hairy BH solutions in a variety of 
theories (no-hair theorems). In theories with non-Abelian fields 
\cite{nonabel}, Einstein-Skyrme \cite{heusler}, and scalar fields with 
non-positive semidefinite potentials \cite{nucsal}, 
BH solutions with hair have been found, 
but the nonlinearity and complexity of the field equations usually demands a 
numerical treatment.

The question that arises is if it is possible to generate new 
static and spherically symmetric (non-vacuum) exact BH solutions 
with matter fields satisfying certain conditions. For instance, 
it is interesting to note that 
when written in the {\it radial gauge}, that is, in the coordinate 
system where the area of the two-spheres is given by $4\pi r^2$, various 
static and spherically symmetric exact BH solutions have the form $g_{tt}=-g_{rr}^{-1}= 
-(1-2m(r)/r)$ where 
$m(r)$ is a rather simple function of $r$ (a power-law like form), 
namely $m_{RN}(r)= M - \frac{Q^2}{2r}$, $m_{SdS}(r)= M + \frac{\Lambda r^3}{6}$,   
for the RN and SdS/SAdS solutions, respectively. 
Then one can ask if it is a mathematical coincidence that in those exact 
solutions, the metric has such a simple form with only one ``degree of freedom'' 
given by $m(r)$. 
A theorem proven below shows that rather than a mathematical coincidence, the 
above form of the metric is a consequence of the features of the 
energy-momentum tensor (hereafter T-tensor) considered. In the above exact solutions, the 
associated T-tensors share some properties that are taken into account 
in the theorem in a general fashion without specifying the nature of the 
matter. Therefore the theorem helps to characterize a whole two-parameter family of 
solutions to the Einstein field equations. The interpretation of the two parameters 
($M$ and $C$) depends 
on the fields associated with the T-tensor and on the boundary conditions 
(one of these parameters is usually associated with the mass). The 
theorem is not restricted to asymptotically flat (AF) spacetimes and can be 
trivially extended as to include a three parameter family of solutions.

\section{The theorem}

{\it Theorem (T1): Let $(M, g_{ab})$ be a four dimensional spacetime 
[sign($g_{ab})= (-,+,+,+)$] such that: (1) It is 
static and spherically symmetric, (2) It satisfies the Einstein field equations, 
(3) In the radial gauge 
(area-$r$) coordinate system adapted to the 
symmetries of the spacetime where $ds^2= -N^2(r) dt^2 + A^2(r) dr^2 + r^2d\theta^2 
+ r^2\sin^2(\theta) \,d\varphi^2$, the energy-momentum tensor $T^{ab}$ satisfies 
the conditions $T^{t}_{\,\,\,t}=T^{r}_{\,\,\,r}$ and $T^{\theta}_{\,\,\,\theta}= 
\kappa T^{r}_{\,\,\,r}$ ($\kappa={\rm const.} \in \mathbb{R}$),
(4) It possess a regular Killing horizon or a regular origin. 
Then, the metric of the spacetime is given by
\begin{equation}
\label{gab}
ds^2= - \left[ 1 - \frac{2m(r)}{r}\right] dt^2 + \left[ 1 - \frac{2m(r)}{r}\right]^{-1} 
dr^2 +  r^2d\theta^2 + r^2\sin^2(\theta)\, d\varphi^2 \,\,\,\,,
\end{equation}
where
\begin{eqnarray}
\label{mass0}
 m(r) &=& \left\{ \begin{array}{lll}
M & \hskip 1cm
 {\rm if} \,\,\,C =0 \\ \\  
 M - \frac{4\pi C r^{2\kappa+1}}{2\kappa +1} & \hskip 1cm
{\rm if} \,\,\,\kappa
\neq -\frac{1}{2} \,\,\,\,{\rm and}\,\,\,C\neq 0 \\ \\
 M - 4\pi C {\rm ln}\left(\frac{r}{|C|}\right) & \hskip 1cm
{\rm if}
\,\,\, \kappa= -\frac{1}{2} \,\,\,\,{\rm and}\,\,\,C\neq 0 
\end{array}\right. 
 \\ \nonumber \\
\label{Tab}
T^{a}_{\,\,\,b} &=& \frac{C}{r^{2(1-\kappa)}}\,{\rm diag}[1,1,\kappa,\kappa]\,\,\,\,,
\end{eqnarray}
$M$ and $C$ are integration constants whose values depend on the 
boundary conditions and on the fundamental constants of the underlying theory 
for the matter fields.}
\bigskip

{\it Proof:} Since the spacetime is assumed to be static and spherically symmetric 
(hypothesis 1) one can always 
adopt a coordinate system adapted to the Killing fields such that the metric $g_{ab}$ 
is given by 
\begin{equation}\label{metric}
ds^2=  -N^2(r) dt^2 + A^2(r) dr^2 + r^2d\theta^2 
+ r^2\sin^2(\theta)\, d\varphi^2\,\,\,.
\end{equation}
Using the Einstein field equations (hypothesis 2)
\begin{equation}\label{Eeqs}
R_{ab} -\frac{1}{2}g_{ab}R = 8\pi T_{ab}\,\,\,,
\end{equation}
it is easily shown that the relevant equations for $N$ and $A$ are
\begin{eqnarray}
\label{Eeq1}
&&
\frac{1}{r^2}\left(A^2-1\right) + \frac{2}{rA}\partial_r A = -8\pi A^2 T^{t}_{\,\,\,t}
\,\,\,\,,\\
\label{Eeq2}
&& \frac{1}{rN}\partial_r N - \frac{1}{r^2}\left(A^2-1\right)- \frac{1}{rA}\partial_r A
= 4\pi A^2 \left(T^{t}_{\,\,\,t} + T^{r}_{\,\,\,r}\right)\,\,\,.
\end{eqnarray}
Moreover, Eq. (\ref{metric}) together with Eq. (\ref{Eeqs}) also imply
\begin{eqnarray}
& & T^{\varphi}_{\,\,\,\varphi} = T^{\theta}_{\,\,\,\theta}\,\,\,,\\
& & T^{ta} = T^{rb} 
= T^{\theta \varphi} =0\,\,\,\,{\rm for}\,\,\,a\neq t\,\,\,\,
{\rm and}\,\,\,b\neq r\,\,\,.
\end{eqnarray}
Therefore
\begin{equation}
\label{T^a_b}
T^{a}_{\,\,\,b}\,\,=
\,\,  \left(
\begin{array}{cccc}
T^{t}_{\,\,\,t}  &0 & 0 & 0 \\ 
         0 & T^{r}_{\,\,\,r} & 0 &0  \\ 
         0 & 0 &T^{\theta}_{\,\,\,\theta} & 0 \\
         0 & 0 & 0 & T^{\theta}_{\,\,\,\theta}   
\end{array}\right) \,\,\,\,.
\end{equation}
Introducing the following reparametrization
\begin{equation}
A(r)= \left( 1 - \frac{2m(r)}{r}\right)^{-1/2}\,\,\,\,,
\end{equation}
Eq.(\ref{Eeq1}) reads
\begin{equation}
\label{Eeq1b}
\partial_r m = -4\pi r^2 T^{t}_{\,\,\,t}\,\,\,.
\end{equation}
Moreover, Eqs. (\ref{Eeq1}) and (\ref{Eeq2}) combine to give
\begin{equation}
\frac{1}{AN}\partial_r(AN)= 4\pi r A^2 \left(T^{r}_{\,\,\,r}- T^{t}_{\,\,\,t}\right)
\,\,\,.
\end{equation}
Now, if $T^{t}_{\,\,\,t}= T^{r}_{\,\,\,r}$ (hypothesis 3), then from the above 
equation on has $N= {\rm const.} A^{-1}$, 
where the constant can be reabsorbed by a redefinition of the time coordinate. So without 
loss of generality 
\begin{equation}\label{g_tt}
N(r)= A^{-1}= \left( 1 - \frac{2m(r)}{r}\right)^{1/2} \,\,\,\,.
\end{equation}
 Therefore, the whole metric is determined by the function 
$m(r)$ which is given from Eq. (\ref{Eeq1b}).

The Einstein Eq. (\ref{Eeqs}) implies the conservation of the energy-momentum tensor
\begin{equation}\label{consT}
\nabla_{b} T^{b}_{\,\,\,a}=0\,\,\,,
\end{equation}
which for the metric Eq. (\ref{metric}) together with Eq. (\ref{g_tt}) 
and the energy-momentum tensor given by Eq. (\ref{T^a_b}) 
provide the equilibrium equation
\begin{equation}\label{equil0}
\partial_rT^{r}_{\,\,\,r} = \left(T^{t}_{\,\,\,t}- T^{r}_{\,\,\,r}\right)\frac{\partial_r N}{N}
-\frac{2}{r}\left(T^{r}_{\,\,\,r}- T^{\theta}_{\,\,\,\theta}\right)\,\,\,.
\end{equation}
Since it is assumed that $T^{t}_{\,\,\,t}= T^{r}_{\,\,\,r}$ and moreover if 
$T^{\theta}_{\,\,\,\theta}= \kappa T^{r}_{\,\,\,r}$ (hypothesis 3), one obtains 
\begin{equation}\label{equil}
\partial_rT^{r}_{\,\,\,r} = 
-\frac{2}{r} T^{r}_{\,\,\,r} (1-\kappa)\,\,\,.
\end{equation}
Therefore, integrating Eq.(\ref{equil}) yields
\begin{equation}\label{T^r_rsol}
T^{r}_{\,\,\,r} = \frac{C}{r^{2(1-\kappa)}} \,\,\,,
\end{equation}
where $C$ is an integration constant. Then, one concludes
\begin{equation}\label{Tab2}
T^{a}_{\,\,\,b} = \frac{C}{r^{2(1-\kappa)}}\,{\rm diag}[1,1,\kappa,\kappa]\,\,\,.
\end{equation}

Finally, using this for $T^{t}_{\,\,\,t}$ in Eq. (\ref{Eeq1b}), one obtains 
\begin{equation}
\label{mass}
 m(r) = \left\{ \begin{array}{lll}
M & \hskip 1cm
 {\rm if} \,\,\,C =0 \\ \\ 
 M - \frac{4\pi C r^{2\kappa+1}}{2\kappa +1} & \hskip 1cm
{\rm if} \,\,\,\kappa
\neq -\frac{1}{2} \,\,\,\,{\rm and}\,\,\,C\neq 0 \\ \\
 M - 4\pi C {\rm ln}\left(\frac{r}{|C|}\right) & \hskip 1cm
{\rm if}
\,\,\, \kappa= -\frac{1}{2} \,\,\,\,{\rm and}\,\,\,C\neq 0 
\end{array}\right. 
\end{equation}

where $M$ is another integration constant. The values of the 
constants $C$ and $M$ depend on the 
boundary conditions and on the fundamental constants of the underlying theory 
for the matter fields. The value of $M$ can be fixed so as to 
avoid naked singularities (hypothesis 4), while a nonzero $C$ together with 
the value of $\kappa$ determine the asymptotic structure of the spacetime. 
Therefore the metric (\ref{metric}) 
together with Eq. (\ref{g_tt}) finally writes
\begin{equation}
ds^2= - \left[ 1 - \frac{2m(r)}{r}\right] dt^2 + \left[ 1 - \frac{2m(r)}{r}\right]^{-1} 
dr^2 +  r^2d\theta^2 + r^2\sin^2(\theta)\, d\varphi^2 \,\,\,\,.
\end{equation}
where $m(r)$ is given by Eq. (\ref{mass}) $\Box$. This concludes the proof.
\bigskip

Table 1 displays some examples of energy-momentum tensors satisfying the conditions 
of the theorem (T1) and which give rise to well known spacetimes. Note that apart from
dS/AdS, no other perfect fluid model verifies the condition 
$T^{t}_{\,\,\,t}=T^{r}_{\,\,\,r}$. Therefore, none of the solutions provided 
by T1 will correspond to more general perfect fluids 
(i.e., inhomogeneous perfect fluids) .

\begin{table}
\begin{tabular}{|l|l|l|c|l|}\hline
{\em Energy-momentum tensor}  
&  {\em Fields} & {\em Spacetime}  & {\em $\kappa$-index} & $C-$parameter \\ \hline\hline
$T_{ab}=0 $ & (vacuum) none & \sch  & $-$  & $C=0$ \\ \hline
$T_{ab}= -\frac{\Lambda}{8\pi} g_{ab}$ & (cosmological constant)   
$\Lambda={\rm const.}$ & de Sitter/anti-de Sitter & $\kappa= 1$ & $C=-\frac{\Lambda}{8\pi}$ \\ \hline
$T_{ab}= (\nabla_a\phi^i) (\nabla_b\phi_i) - \frac{g_{ab}}{2}(\nabla\phi^i)^2$ & (global monopole)
 $\phi^i= \eta x^i/r$ & Black Hole with a & $\kappa= 0$ & $C=-\eta^2$ \\ 
$\,\,\,\,\,\,\,\,\,\,\,\,\,-g_{ab}\frac{\lambda}{4}\left(\phi^i\phi_i -\eta^2\right)^2$ 
& & global monopole & & \\\hline
$T_{ab}=\frac{1}{4\pi}\left( F_{ac}F_b^{\,\,c}-\frac{1}{4}g_{ab}F_{cd}F^{cd}
\right)$ & (electric field) $A_a= -\delta^t_a \frac{Q}{r}$ & 
 Reissner-Nordstrom  & $\kappa= -1$ & $C=-\frac{Q^2}{8\pi}$  \\ \hline
\end{tabular}
\caption{Examples of energy-momentum tensors satisfying the conditions 
of the theorem (T1), and 
the corresponding static and spherically symmetric spacetimes they generate. The table 
shows the values $\kappa$ and $C$ associated with each matter model.}
\end{table}

The hypothesis 
(4) concerning the existence of a Killing horizon or a regular origin 
was not required a priori for the formal proof of T1. However, 
the assumption $T^{t}_{\,\,\,t}= T^{r}_{\,\,\,r}$ is indeed suggested by
the general condition of a regular horizon in these coordinates 
(i.e., regardless of the specific form of the energy-momentum tensor in these coordinates): 
if one demands that the components 
$T^{r}_{\,\,\,r}$, $T^{\theta}_{\,\,\,\theta}$ and $\partial_r T^{r}_{\,\,\,r}$ 
be bounded at the horizon $r_h$ 
[$N(r_h)=0$], then from Eq. (\ref{equil0}), one obtains the regularity condition 
$\left(T^{t}_{\,\,\,t}- T^{r}_{\,\,\,r}\right)|_{r_h}=0$. 
Note also that if regularity at the origin is imposed, then 
$\left(T^{r}_{\,\,\,r}- T^{\theta}_{\,\,\,\theta}\right)|_{r=0}=0$. If both 
conditions can be imposed, then the solution 
suggests the existence of a globally regular black hole. 
The well known example 
satisfying these conditions is 
the (pure) dS solution (i.e., with $M=0$) 
which possesses a cosmological horizon with a regular origin and 
which corresponds to $\kappa=1$. The T-tensor indeed verifies (see Table 1) 
$T^{t}_{\,\,\,t}= T^{r}_{\,\,\,r}= T^{\theta}_{\,\,\,\theta}= 
T^{\varphi}_{\,\,\,\varphi}={\rm const.}$, and so the regularity conditions at the 
horizon and origin are automatically satisfied. 
  
The globally regular AF exact BH solution with a
nonlinear electrodynamics source \cite{eloy0}, shows that 
$T^{t}_{\,\,\,t}= T^{r}_{\,\,\,r}$. In particular, 
the equality holds at the horizon. The solution has therefore the form $N=A^{-1}$. 
Nonetheless, 
the condition 
$T^{\theta}_{\,\,\,\theta}= T^{r}_{\,\,\,r} $ 
is not verified by the solution in general although it is satisfied at the origin. Since 
in that solution $T^{\theta}_{\,\,\,\theta}\neq \kappa T^{r}_{\,\,\,r}$ (for all $r$) 
then the theorem T1 is not able to account for it. In fact, one can associate a mass function 
$m(r)$ with that solution, but the expression is not as simple as in Eq. (\ref{mass}). 
Hence, it is important to stress that T1 does not cover all possible static and spherically symmetric 
exact BH solutions, not even all those 
where $N=A^{-1}$ ($T^{t}_{\,\,\,t}= T^{r}_{\,\,\,r}$), as it was just illustrated 
for $T^{\theta}_{\,\,\,\theta} \neq \kappa T^{r}_{\,\,\,r} $ \cite{eloy0}.

Now, if the condition $T^{\theta}_{\,\,\,\theta}= 
\kappa T^{r}_{\,\,\,r}$ is dropped, the metric has still the form as proposed in 
T1 but $m(r)$ is not given explicitly by Eq. (\ref{mass}), but rather by 
integrating Eq. (\ref{Eeq1b}). Moreover, in order to integrate the equation of 
equilibrium
\begin{equation}\label{equil2}
\partial_rT^{r}_{\,\,\,r} = 
-\frac{2}{r}\left(T^{r}_{\,\,\,r}- T^{\theta}_{\,\,\,\theta}\right)\,\,\,.
\end{equation}
a closure relationship between $T^{r}_{\,\,\,r}$ and $T^{\theta}_{\,\,\,\theta}$ is 
needed. An example of this is the Reissner-Nordstrom-de Sitter solution (RNdS) 
\cite{Kramer}:
\be\label{SdSRN}
m(r)= M - \frac{Q^2}{2r} + \frac{\Lambda r^3}{6}  \,\,\,.
\end{equation}
In order to recover this result, one needs to consider 
the T-tensor as the sum of the electromagnetic and the cosmological-constant 
contributions:
\begin{equation}
T^{a}_{\,\,\,b} = T^{a}_{EM\,\,\,b} + T^{a}_{\Lambda\,\,\,b} \,\,\,.
\end{equation}
Now, clearly $T^{t}_{\,\,\,t}=T^{r}_{\,\,\,r}$ since the individual contributions 
verify this. On the other hand $T^{\theta}_{\,\,\,\theta} \neq \kappa T^{r}_{\,\,\,r} $. In fact,  
adding the above expressions for $T^{r}_{\,\,\,r}$ and $T^{\theta}_{\,\,\,\theta}$, 
and using $T^{\theta}_{EM\,\,\,\theta} = - T^{r}_{EM\,\,\,r}$ 
and $T^{\theta}_{\Lambda\,\,\,\theta}= 
T^{r}_{\Lambda\,\,\,r}= -\Lambda/(8\pi)$, it turns out
\be\label{newcond}
T^{\theta}_{\,\,\,\theta}= - T^{r}_{\,\,\,r} - \frac{\Lambda}{4\pi}\,\,\,.
\end{equation}
In this way Eq. (\ref{equil2}) can be integrated explicitly as
\be\label{TSdSRN}
T^{r}_{\,\,\,r}= Cr^{-4} - \frac{\Lambda}{8\pi}= T^{t}_{\,\,\,t} \,\,\,,
\end{equation}
and in turn Eq. (\ref{newcond}) becomes
\be\label{newcond2}
T^{\theta}_{\,\,\,\theta}= -  Cr^{-4} - \frac{\Lambda}{8\pi}\,\,\,.
\end{equation}
These results are obvious in light that $T^{ab}_{EM}$ and $T^{ab}_{\Lambda}$ 
conserves separately and therefore the total solution for $T^{ab}$ is the solution 
for $T^{ab}_{EM}$ plus the one for $T^{ab}_{\Lambda}$.

Since $T^{t}_{\,\,\,t}= T^{r}_{\,\,\,r}$, Eq. (\ref{Eeq1b}) integrates as follows:
\be\label{mSdSRN}
m(r)= M -\frac{Q^2}{2r} + \frac{\Lambda r^3}{6} \,\,\,,
\end{equation}
where $C= -Q^2/(8\pi)$. This is a three-parameter 
class of spacetime characterized by $M,Q$ and $\Lambda$. This shows, that 
T1 can be easily generalized as to include this case (hereafter referred to 
as the trivial extension of the theorem T1) by 
replacing the condition $T^{\theta}_{\,\,\,\theta}= 
\kappa T^{r}_{\,\,\,r}$ with 
\be\label{newcond0}
T^{\theta}_{\,\,\,\theta}= \kappa T^{r}_{\,\,\,r} - \frac{\Lambda (1-\kappa)}{8\pi}
\,\,\,.
\end{equation}
For $\kappa = 1$ the latter condition reduces to the case of dS/AdS of T1. As is 
shown below, the new condition is equivalent to say that $T^{a}_{\,\,\,b}= 
T^{a}_{{\rm fields}\,\,\,b} - \frac{\Lambda}{8\pi} \delta^{a}_{\,\,\,b}$ with 
$T^{a}_{{\rm fields}\,\,\,b}$ satisfying the conditions of T1, 
that is, $T^{t}_{{\rm fields}\,\,\,t}=T^{r}_{{\rm fields}\,\,\,r}$ and 
$T^{\theta}_{{\rm fields}\,\,\,\theta}= \kappa T^{r}_{{\rm fields}\,\,\,r}$. 
The spacetime is then parametrized by $M$, $C$ and $\Lambda$. 
Indeed, using Eq. (\ref{newcond0}) in Eq. (\ref{equil2}) one obtains 
\be\label{T^r_rsolext}
T^{r}_{\,\,\,r}= \frac{C}{r^{2(1-\kappa)}} - \frac{\Lambda}{8\pi} 
= T^{r}_{{\rm fields}\,\,\,r} - \frac{\Lambda}{8\pi}\,\,\,,
\end{equation}
and in turn using this result in Eq. (\ref{newcond0}) one has
\be
T^{\theta}_{\,\,\,\theta}= \frac{\kappa C}{r^{2(1-\kappa)}} - \frac{\Lambda}{8\pi}
= \kappa T^{r}_{{\rm fields}\,\,\,r} - \frac{\Lambda}{8\pi} \,\,\,,
\end{equation}
which confirms the announced result. Finally, $T^{t}_{\,\,\,t}= T^{r}_{\,\,\,r}$ 
since one assumes that $T^{t}_{{\rm fields}\,\,\,t}= T^{r}_{{\rm fields}\,\,\,r}$, 
then using (\ref{T^r_rsolext}) in Eq. (\ref{Eeq1b}) one can straightforwardly 
integrate for $m(r)$. Then the trivial generalization of T1 
writes in full detail as follows:

{\it Theorem (T2): Let $(M, g_{ab})$ be a four dimensional spacetime 
[sign($g_{ab})= (-,+,+,+)$] such that: (1) It is 
static and spherically symmetric, (2) It satisfies the Einstein field equations, 
(3) The total energy-momentum tensor $T^{a}_{\,\,\,b}$ is given by  
$T^{a}_{\,\,\,b}= T^{a}_{{\rm fields}\,\,\,b} - \frac{\Lambda}{8\pi} \delta^a_{\,\,\,b}$, 
where $\Lambda$ is a cosmological constant and 
$T^a_{{\rm fields}\,\,\,b}$ is the energy-momentum tensor of the matter 
fields, (4) In the radial gauge 
(area-$r$) coordinate system adapted to the 
symmetries of the spacetime where $ds^2= -N^2(r) dt^2 + A^2(r) dr^2 + r^2d\theta^2 
+ r^2\sin^2(\theta)\, d\varphi^2$, the energy-momentum tensor 
$T^a_{{\rm fields}\,\,\,b}$ satisfies 
the conditions $T^{t}_{{\rm fields}\,\,\,t}=T^{r}_{{\rm fields}\,\,\,r}$ 
and $T^{\theta}_{{\rm fields}\,\,\,\theta}= 
\kappa T^{r}_{{\rm fields}\,\,\,r}$ ($\kappa={\rm const.} \in \mathbb{R}$),
(5) It possess a regular Killing horizon or a regular origin. 
Then, the metric of the spacetime is given by}
\begin{equation}
\label{gabext}
ds^2= - \left[ 1 - \frac{2m(r)}{r}\right] dt^2 + \left[ 1 - \frac{2m(r)}{r}\right]^{-1} 
dr^2 +  r^2d\theta^2 + r^2\sin^2(\theta)\, d\varphi^2 \,\,\,\,,
\end{equation}
where
\begin{eqnarray}
\label{mass0ext}
 m(r) &=& \left\{ \begin{array}{lll}
M + \frac{\Lambda  r^3}{6}  & \hskip 1cm
 {\rm if} \,\,\,C =0 \\ \\  
 M - \frac{4\pi C r^{2\kappa+1}}{2\kappa +1} + \frac{\Lambda  r^3}{6}  & \hskip 1cm
{\rm if} \,\,\,\kappa
\neq -\frac{1}{2} \,\,\,\,{\rm and}\,\,\,C\neq 0 \\ \\
 M - 4\pi C {\rm ln}\left(\frac{r}{|C|}\right) + \frac{\Lambda  r^3}{6}  & \hskip 1cm
{\rm if}
\,\,\, \kappa= -\frac{1}{2} \,\,\,\,{\rm and}\,\,\,C\neq 0 
\end{array}\right. 
 \\ \nonumber \\
\label{Tabext}
T^{a}_{\,\,\,b} &=& \frac{C}{r^{2(1-\kappa)}}\,{\rm diag}[1,1,\kappa,\kappa]
- \frac{\Lambda}{8\pi} \,{\rm diag}[1,1,1,1] \,\,\,\,.
\end{eqnarray}
It is to note that the dS/AdS spacetime given by Eqs. (\ref{mass0}) and 
(\ref{Tab}) with $\kappa=1$ and $C= - \frac{\Lambda}{8\pi}$, corresponds 
in Eqs. (\ref{mass0ext}) and (\ref{Tabext}) (which {\it ab-initio} includes the cosmological 
constant) to $C=0$. When $\Lambda =0$, T2 reduces to 
T1 \cite{comment0}. Theorem T2 provides thus static and spherically symmetric exact 
solutions to the Einstein field equations with matter fields satisfying the conditions 
of T1 and in the presence of a cosmological constant. 
\bigskip

In more complicated situations, where the relationship between $T^{t}_{\,\,\,t}$, 
$T^{r}_{\,\,\,r}$ and $T^{\theta}_{\,\,\,\theta}$ is not as simple as in T1 
(or its trivial extension T2), the usual procedure to find 
solutions is not to use the first order differential equation for $T^{r}_{\,\,\,r}$ 
Eq. (\ref{equil0}), but rather to employ the equation of motion 
for the matter fields themselves (i.e., the fundamental fields 
entering in the energy-momentum tensor) which results from Eq. (\ref{equil0}). It is 
a higher order equation for the fields (usually a second order differential equation). 
This is what happens in most of the hairy BH solutions where the relationship 
between the components of the T-tensor is not linear and therefore one usually 
requires a numerical analysis to solve the equations. On the other hand, as showed in Sec. V, 
in some cases one can 
push the equations as far as possible so that T1 or T2 applies. This is 
rather surprising an allows one to understand the existence of 
exact BH solutions in more complicated theories of gravity.

For simplicity, only the results of theorem T1 will be discussed 
in sections III and IV, since for the trivial extension T2 
the analogous results are easily obtained.

\section{Asymptotics and regularity}
Scalar invariants are often used to test the regularity at the horizons and to 
analyze the falloff of the geometry asymptotically. Among these, the 
Ricci scalar, the squared Ricci, and the Kretschmann invariant (for more scalars 
see Ref. \cite{scalars}). For the solution given by T1, 
the expressions for these scalars are as follows:
\begin{eqnarray}
  & R &= -8\pi T^{a}_{\,\,\,a}= \frac{4(1+\kappa)\partial_r m}{r^2}=
-\frac{16\pi C(\kappa +1)}{r^{2(1-\kappa)}}  \,\,\,,\\
 &R^{ab} R_{ab}& = (8\pi)^2 T^{ab} T_{ab} = \frac{8(1+\kappa^2)(\partial_r m)^2}{r^4}
= \frac{128\pi^2 C^2 (1+\kappa^2)}{r^{4(1-\kappa)}}\,\,\,,\\
\label{kret1}
 &{\cal K} & = R^{abcd}R_{abcd}= \frac{4}{r^6}\left\{
8r^2 (\partial_r m)^2  - 16rm (\partial_r m) + 12m^2 + r^4(\partial^2_{rr} m)^2
-4r^3 (\partial^2_{rr} m) (\partial_r m) + 4mr^2 \partial^2_{rr} m\right\} \nonumber \\
& & = \frac{48M^2}{r^6} + 
\frac{256\pi^2 C^2}{r^6}\left(\frac{r^{2\kappa +1}}{2\kappa +1}\right)^2 
\left[\left(\kappa-1\right)^2 \left(2\kappa + 1\right)^2 + 2\kappa\left(4\kappa-1\right)\right]
\nonumber \\
&&\,\,\,\, -\frac{16\pi M C}{r^6} \frac{r^{2\kappa +1}}{2\kappa +1}\left[\left(4\kappa-3\right)^2 -1\right]\,\,\,\,,\,\,\,{\rm if}\,\,\,\kappa\neq -\frac{1}{2} \,\,\,,\\
\label{kret2}
& & =  \frac{48M^2}{r^6} + \frac{64\pi C}{r^6}
\left\{13\pi C + M\left[5- 6{\rm ln}\left(\frac{r}{|C|}\right)\right]
-4\pi C \left[5- 3{\rm ln}\left(\frac{r}{|C|}\right)\right] 
{\rm ln}\left(\frac{r}{|C|}\right) \right\} \,\,\,,\nonumber \\
&&\hskip .5cm {\rm if}\,\,\,\,\,\kappa = -\frac{1}{2}\,\,\,{\rm and}\,\,\,C\neq 0 \,\,\,.
\end{eqnarray}

For $C=0$ the above invariants reduce to the \sch case. 
These invariants are regular everywhere except at the origin $r=0$ where 
they can become singular. 

The null Killing horizons are located at $r=r_h$ satisfying
\begin{equation}\label{horizon}
g_{tt}(r_h) =-N^2(r_h)=0= 
\left\{ \begin{array}{lll} 
 1-\frac{2M}{r_h} & \hskip 1cm
 {\rm if} \,\,\,C =0 \\ \\
 1-\frac{2M}{r_h} + \frac{8\pi C {r_h}^{2\kappa}}{2\kappa +1} & \hskip 1cm
{\rm if} \,\,\,\kappa
\neq -\frac{1}{2} \,\,\,\,,\,\,\,C\neq 0 \\ \\
 1-\frac{2M}{r_h} + \frac{8\pi C}{r_h} {\rm ln}\left(\frac{r_h}{|C|}\right) & \hskip 1cm
{\rm if}
\,\,\, \kappa= -\frac{1}{2} \,\,\,\,,\,\,\,C\neq 0 
\end{array}\right. 
\end{equation}

{\bf Case $\kappa\geq 1$}. The spacetime is not AF (unless it is the trivial Minkowski 
spacetime $M=0=C$) and can be 
singular at $r=0$ but it is regular everywhere else. 
Regularity at the origin requires $M=0$ [cf. Eq. (\ref{kret1})]. 
From Eq. (\ref{T^r_rsol}) one can also see 
that $T^{r}_{\,\,\,r}= T^{\theta}_{\,\,\,\theta}$ at the origin regardless of 
the value $\kappa$, which 
is consistent what it was said previously in Sec. II after proving the theorem T1. 
The asymptotic 
structure depends on the value of $\kappa$ and the sign of $C$. 
For $C>0$ there can be an horizon only if $M\neq 0$ [cf. Eq. (\ref{horizon}) ]. 
For $C<0$ a cosmological and event horizon may be present.
This case includes $\kappa=1$ which is associated with a cosmological constant 
(cf. Table 1) and the spacetime is asymptotically dS/AdS depending on the sign of $C$. 
Asymptotically the invariants behave as
\begin{eqnarray}
R &\sim& -C r^{2(\kappa-1)}  \,\,\,,\\
R^{ab} R_{ab} &\sim& C^2 r^{4(\kappa-1)} \,\,\,\,, \\
{\cal K} &\sim& C^2 r^{4(\kappa-1)}  \,\,\,\,.
\end{eqnarray}

\bigskip

{\bf Case $\kappa< 1$}. In this case, a nontrivial spacetime cannot be regular at the 
origin. However, a regular horizon can be present covering the singularity at $r=0$. 
Several horizons appear depending on the values of $\kappa$, $M$ and $C$. The following 
subcases provide more details.
\bigskip

{\bf Case $-1/2\leq \kappa<1$}. The spacetime is not AF 
(except in vacuum $C=0$). Then, the 
Arnowitt-Deser-Misner mass $M_{ADM}$ (or alternatively the Komar mass) 
given simply by $m(\infty)$ is not well defined. 
In such cases, another definition of $M_{ADM}$ has to be provided according to the 
asymptotic structure of the corresponding spacetime. For instance, the case $\kappa=0$ 
will be discussed in Sec V (see also Table 1). 
The asymptotic behavior of the invariants is given by
\begin{equation}
\begin{array}{ll} 
 R \sim -C r^{-2(1-\kappa)} & \hskip 1cm
{\rm if}\,\,\,-1/2\leq \kappa<1  \,\,\,,\\ \\
R^{ab} R_{ab} \sim C^2 r^{-4(1-\kappa)} & \hskip 1cm
{\rm if}\,\,\,\,\,-1/2\leq \kappa<1\,\,\,\,, \\ \\
 {\cal K}\sim C^2 r^{-4(1-\kappa)} & \hskip 1cm
{\rm if}\,\,\,-1/2<\kappa< 1 \,\,\,\,,\\ \\
{\cal K}\sim \left[C r^{-3}{\rm ln}\left(\frac{r}{|C|}\right) \right]^2 & \hskip 1cm
{\rm if}\,\,\,\kappa=-1/2\,\,\,\,,\,\,\,\,C\neq 0
\end{array}
\end{equation}
For $-1/2< \kappa<1$ and $C> 0$ one requires that $M\neq 0$ to avoid a naked 
singularity [cf. Eq. (\ref{horizon}) ]. If $C<0$ and $M=0$, $\kappa \neq -1/2$ , 
then $r_h= \left[\frac{2\kappa +1}{8\pi |C|}\right]^{\frac{1}{2\kappa}}$. 
If $\kappa=-1/2$ some horizons may be present as well even with 
$M=0$, and they are located at the roots of $x + 8\pi\, {\rm sign}(C) {\rm ln}x=0$ 
(where $x= r_h/|C|$ and ${\rm sign}(C)= C/|C|$). 
For $C>0$, $r_h \approx 0.96243\, C$, and for $C<0$ there are 
two horizons: $r_h^- \approx 1.04235 \, |C| $ and 
$r_h^+ \approx 120.408\, |C| $. 
\bigskip

{\bf Case $\kappa <-1/2$.} The spacetime is AF and the horizons can appear 
at different $r$'s as well [cf. Eq. (\ref{horizon}) ]. The ADM-mass is given by $M_{ADM}=M$.
The value $\kappa=-1$ give rise to the RN solution with $C= -Q^2/8\pi$ (cf. Table 1). The asymptotic behavior of the invariants is
\begin{eqnarray}
R &\sim& -C (1-|\kappa|) r^{-2(1 + |\kappa|)}  \,\,\,,\\
R^{ab} R_{ab} &\sim& C^2 r^{-4(1+|\kappa|)} \,\,\,, \\
{\cal K} &\sim& \frac{48M^2}{r^6}  \,\,\,.
\end{eqnarray}

\section{Energy conditions and causality}
In addition to the different asymptotic structures that can arise from the values 
of $\kappa$, the matter associated with those spacetimes might 
violate the energy conditions (EC). In order to discuss such EC of the matter fields, 
it is useful to remind that in 
the rest frame associated to the time-like Killing vector $(\partial/\partial t)^a$, 
the energy-density of the matter will be given by [cf. Eq. (\ref{Tab2}) ],
\begin{equation}
\rho= -T^{t}_{\,\,\,t}= -T^{r}_{\,\,\,r}=  -\frac{C}{r^{2(1-\kappa)}}\,\,\,,
\end{equation}
and the principal pressures by
\begin{equation}
p_i= T^{i}_{\,\,\,i} \,\,({\rm no \,\,sum\,\,\, convention})\,\,\,.
\end{equation}
Therefore $p_r= T^{r}_{\,\,\,r}= -\rho$ and $p_{\theta}=p_{\phi}=T^{\theta}_{\,\,\,\theta}= 
\kappa p_r= -\kappa \rho $ (hypothesis 3 of T1).
One can introduce the effective pressure (in the same rest-frame) by averaging the principal 
pressures:
\begin{equation}
p_{\it eff}:= \frac{1}{3}\sum_{i=1}^{3}T^{i}_{\,\,\,i}= (1+2\kappa) \frac{T^{r}_{\,\,\,r}}{3} \,\,\,.
\end{equation}
Therefore from the above definitions one can write an effective 
``equation of state'' as follows
\begin{equation}
p_{\it eff}= - (1+2\kappa)\frac{\rho}{3} \,\,\,\,.
\end{equation}
So, for the cases $\kappa=1$ and $\kappa=-1$ corresponding to the energy-momentum tensors 
associated with a cosmological constant and the electromagnetic field respectively, 
one recovers the usual equations of state $p_{\it eff}=- \rho$ and $p_{\it eff}= \rho/3$, 
respectively. 

Since $p_{\it eff}$ includes the tangential pressures, which for $\kappa\neq1$ are 
not equal to the radial pressure, then the case $\kappa=-1/2$ leading to 
$p_{\it eff}=0$, does not pertain to a dust pressureless fluid, since 
there exists indeed a radial pressure. Now, according to the weak energy condition (WEC) 
\cite{wald},
\begin{equation}
\rho\geq0 \,\,\,\,\,{\rm and}\,\,\,\rho + p_i \geq 0 \,\,\, (i=1,2,3)\,\,\,.
\end{equation}
Therefore the WEC is satisfied if $C\leq 0$ and $\kappa\leq 1$.

On the other hand, the strong energy condition \cite{wald},
\begin{equation}
\rho + \sum_{i=1}^{3}p_i \geq0 \,\,\,\,\,{\rm and}\,\,\,\rho + p_i \geq 0 \,\,\, 
(i=1,2,3)\,\,\,\,,
\end{equation}
will be verified if $C\leq 0$ and $\kappa \leq 0$ or if $C\geq 0$ and 
$\kappa \geq 1$ (cf. with AdS solution).
 
Finally, the dominant energy condition \cite{wald},
\begin{equation}
\rho \geq |p_i|  \,\,\, (i=1,2,3) \,\,\,\,,
\end{equation}
 is satisfied if $C\leq 0$ and $|\kappa|\leq 1 $. Therefore the three EC are compatible if $C\leq 0$ and 
$-1\leq \kappa \leq 0$. Moreover, if $-1\leq \kappa \leq -1/2$, all the energy conditions 
are verified, and furthermore, the effective pressure $p_{\it eff}\geq 0$ 
and the causality condition 
(i.e., the mean speed of sound in the medium less than the speed of light) 
$0 \leq p_{\it eff}/\rho < 1$ will be satisfied as well.

\section{Applications and discussion of the theorem}
Clearly, the theorem T1 (or its trivial extension T2) just proved, includes a broad class of T-tensors, some of which 
might be unphysical or  
do not represent plausible matter fields \cite{comment}. However, a priori this is difficult 
to judge only by the conditions imposed on $T^{ab}$. For instance, the case $\kappa=0$, 
corresponding to $T^{\theta}_{\,\,\,\theta}=T^{\varphi}_{\,\,\,\varphi}=0$ but not 
necessarily $T^{r}_{\,\,\,r}=0$, could seem rather odd and not to 
correspond to a physical situation at first glance (cf. Table 1). Nonetheless, the T-tensor that indeed verifies 
those conditions and furthermore $T^{t}_{\,\,\,t}= T^{r}_{\,\,\,r}\neq 0$ 
represents the matter distribution of a trivial global monopole 
(see Ref. \cite{vilenkin} for a review on the subject) $\phi^i= \eta x^i/r$ 
($i=1,2,3$) that remains in the true vacuum of its potential: 
$V(\phi^i\phi_i)=\frac{\lambda}{4}\left(\phi^i\phi_i -\eta^2\right)^2\equiv0$, and which is given by
\begin{equation}
T_{ab}= (\nabla_a\phi^i)(\nabla_b\phi_i)
- g_{ab}\left[\frac{1}{2}(\nabla\phi^i)^2 + V(\phi^i\phi_i)\right]
= (\nabla_a\phi^i) (\nabla_b\phi_i) -\frac{g_{ab}}{2}(\nabla\phi^i)^2
\,\,\,\,.
\end{equation}
In particular $T^{r}_{\,\,\,r}= T^{t}_{\,\,\,t}= -\eta^2/r^2$, which fixes the value 
$C=-\eta^2$ in Eq. (\ref{T^r_rsol}). 
The spacetime generated by this kind of source is the one of a static and spherically 
symmetric black hole with a global monopole inside 
\cite{nucsud0}. The spacetime is not AF but AF with a solid angle deficit 
given by $\Delta= 8\pi \eta^2$ ($ 0 \leq \Delta <1$). The corresponding ADM-mass 
has been well defined \cite{nucsud1}, and it is indeed given 
by $M_{\rm ADM_\Delta}= M(1-\Delta)^{-3/2}$ where $M\propto \eta^{-1}\lambda^{-1/2}$ 
\cite{commentmon}. 
This example shows, that as odd as it could appear, different values of $\kappa$ could 
indeed be associated with T-tensors interesting from the mathematical or the 
theoretical-physics point of view. 

Another natural question that arises is: Until what extent T1 or T2 of Sec. II covers 
the maximum of situations ?, or in which situations the hypothesis need to be 
strengthen or relaxed ?

 In order to give some insight relevant to this question, more complicated 
examples corresponding to scalar-tensor theories (STT) of gravity are considered below. 
The general action for a STT with a single
scalar field and a cosmological constant is given by 
\begin{equation}
  \label{jordan}
S[g_{ab}, \phi] = \int \left\{ \frac{1}{16\pi} F(\phi) R 
-\frac{\Lambda}{8\pi}
-\left( \frac{1}{2}(\nabla \phi)^2 + V(\phi) \right) \right\} \sqrt{-g} d^4x \,\,\,.
\end{equation}
 The equations of motion obtained from the above action are
\begin{eqnarray}  
G_{ab} &=& 8\pi T_{ab}\,\,\,\,, \\
\label{effTmunu}
T_{ab} &=& G_{eff}\left( T_{ab}^F + T_{ab}^{\phi} + T_{ab}^{\Lambda}\right)\,\,\,\,, \\
T_{ab}^F &= & \frac{1}{8\pi}\left[\nabla_a\left(\partial_\phi
F\nabla_b\phi\right) - g_{ab}\nabla_c \left(\partial_\phi
F\nabla^c \phi\right)\right] \,\,\,\,, \\
T_{ab}^{{\rm \phi}} &= & (\nabla_a \phi)(\nabla_b \phi) - g_{ab}
\left[ \frac{1}{2}(\nabla \phi)^2 + V(\phi)\right ] \,\,\,\,, \\
T_{ab}^{\Lambda} &=& -\frac{\Lambda}{8\pi}g_{ab} \,\,\,,\\
\label{Geff}
G_{eff} &:=& \frac{1}{F} \,\,\,\,, 
\end{eqnarray}
\begin{widetext}
\begin{equation}
\label{KGgen}
{\Box \phi} = \frac{ F\partial_\phi V- 2(\partial_\phi F) V -\frac{1}{2}
(\partial_\phi F) \left( 1 + \frac{3\partial^2_{\phi\phi} F}{8\pi}
\right)(\nabla \phi)^2 + \frac{1}{2}(\partial_\phi F) T_{\Lambda} }{ F + 
\frac{ 3(\partial_\phi F)^2}{16\pi} }\,\,\,\,.
\end{equation}
\end{widetext}
where $T_{\Lambda}$ stands for the trace of $T^{ab}_{\Lambda}$ and 
$G_{ab}=R_{ab} -\frac{1}{2}g_{ab}R$.

Two particular examples of the above STT are considered below.

The first example is the 
Bekenstein-Bocharova-Bronnikov-Melnikov (BBBM) asymptotically flat BH solution
of a scalar field conformally coupled to the curvature \cite{Bek1,BBM}, 
and which is identified with the 
extreme Reissner-Nordstrom solution. Therefore in principle this 
solution should be covered by T1. The solution has been the object 
of controversy and I will elaborate more on this once the solution is recovered 
along the lines of the theorem T1.

The BBBM solution corresponds to the choice
\begin{equation}\label{F}
F(\phi)= 1 + 16\pi \xi \phi^2 = 1 - \frac{4\pi}{3} \phi^2\,\,\,,
\end{equation}
with $\xi=-1/12$ and $V(\phi)=0= \Lambda$. The theory is invariant 
with respect to the conformal transformations:
$g_{ab}\rightarrow \Omega^2 g_{ab}$, $\phi\rightarrow \Omega^{-1} \phi$. From 
Eqs. (\ref{effTmunu})$-$(\ref{Geff}), one writes the following effective energy-momentum tensor
\begin{equation}
\label{TEC}
T_{ab}= \frac{1}{6F}\left[
4(\nabla_a\phi)(\nabla_b\phi) - 2\phi \nabla_a\nabla_b\phi - 
g_{ab}(\nabla^c\phi) (\nabla_c\phi)\right] \,\,\,.
\end{equation} 
Like in the electromagnetic case, the T-tensor is traceless $T^a_{\,\,\,a}=0$. 
Moreover, for the choice of $F$ Eq. (\ref{F}) and with $V(\phi)=0= \Lambda$, 
Eq. (\ref{KGgen}) reduces simply
\begin{equation}\label{KG}
\Box \phi =0 \,\,\,\,.
\end{equation}
From the form of $T_{ab}$, it is clear 
that T1 cannot a priori be applied. However, 
the conditions required on $T_{ab}$ will be imposed ``by force'' and then check that 
all the equations are fully consistent [except perhaps for one subtle point that 
is discussed below Eq. (\ref{phi3})]. 
First, for the metric Eq. (\ref{metric}), the Klein-Gordon Eq. (\ref{KG}) 
reads
 \begin{equation}\label{KG2}
\frac{1}{r^2 A N}\partial_r\left(r^2\frac{N}{A}\partial_r\phi\right)=0\,\,\,\,.
\end{equation}
In addition, from Eq. (\ref{TEC}) 
a simple calculation that uses Eq. (\ref{KG2}) to replace 
the second order derivatives, leads to
\begin{equation}
\label{Treg}
T^{r}_{\,\,\,r}- T^{t}_{\,\,\,t}= \frac{2(\partial_r\phi)}{3FA^2} \frac{1}{rN}
\partial_r\left(rN\phi\right)\,\,\,.
\end{equation}
Now, for $T^{r}_{\,\,\,r}= T^{t}_{\,\,\,t}$, one requires  
\begin{equation}\label{phi0}
rN\phi= {\rm const.}= d\,\,\,.
\end{equation}
In that case $A= N^{-1}$, and therefore
\begin{equation}\label{phi1}
\phi= \frac{dA}{r}\,\,\,.
\end{equation}
On the other hand Eq. (\ref{KG2}) leads to
\begin{equation}\label{phip}
\partial_r\phi= \frac{a A^2}{r^2}\,\,,\,\,\,a={\rm const.}
\end{equation}
Therefore, Eqs. (\ref{phi1}) and (\ref{phip}) imply
\begin{equation}\label{phip2}
\partial_r\phi= \frac{1}{e}\phi^2\,\,\,.
\end{equation}
where $e=d^2/a={\rm const.}$ The solution of this equation is given by
\begin{equation}\label{phi2}
\phi= -\frac{e}{r-M}\,\,\,,
\end{equation}
 where $M$ is an integration constant.
Now, from Eq. (\ref{TEC}), it turns out
\begin{equation}
T^{t}_{\,\,\,t}= \frac{d^2}{6eFr^2} \left(\partial_r\phi + \frac{2\phi}{r}\right)\,\,\,.
\end{equation}
where Eqs. (\ref{phi1}) and (\ref{phip}) were used. Finally, using Eqs. (\ref{F}) and 
(\ref{phi2}), one has
\begin{equation}\label{Tttbek0}
T^{t}_{\,\,\,t}= - \frac{d^2 \left( 1 -\frac{2M}{r}\right)}{6r^4
\left[ 1 -\frac{2M}{r} + \frac{M^2}{r^2}\left(1 - \frac{4\pi e^2}{3M^2}\right)
\right]}\,\,\,.
\end{equation}
In the same way
\begin{equation}
T^{\theta}_{\,\,\,\theta}= \frac{e^2 \left( 1 -\frac{2M}{r}\right)}{6r^4
\left[ 1 -\frac{2M}{r} + \frac{M^2}{r^2}\left(1 - \frac{4\pi e^2}{3M^2}\right)
\right]}\,\,\,.
\end{equation}
Clearly the choice $d^2=e^2$ leads to the requirement 
$T^{\theta}_{\,\,\,\theta}= -T^{t}_{\,\,\,t}= -T^{r}_{\,\,\,r}  $. 
To conclude, using Eqs. (\ref{phi0}) and (\ref{phi2}), one is led to
\begin{eqnarray}
N &=& A^{-1}= 1 -\frac{M}{r} = \left( 1 -\frac{2m(r)}{r}\right)^{1/2}\,\,\,,\\
\label{mbek}
m(r)&=& M - \frac{M^2}{2r}\,\,\,\,.
\end{eqnarray}
One can check independently that Eq. (\ref{mbek}) is consistent with Eqs. (\ref{Eeq1b}) 
and (\ref{Tttbek0}) if and only if $d^2=e^2= 3M^2/(4\pi)$ for which
\begin{equation}\label{Tttbek}
T^{t}_{\,\,\,t}=  - \frac{M^2}{8\pi r^4}= T^{r}_{\,\,\,r}
\,\,\,.
\end{equation}
In this way, one recovers the result of T1 with 
$\kappa=-1$, and $C= -M^2/(8\pi)$ which is no other but the extreme RN 
spacetime with the scalar field solution given by Eq. (\ref{phi2}). In summary, the 
black hole solution found by BBBM is \cite{comment2}
\begin{eqnarray}
\label{solBBBM}
ds^2 &=& -\left( 1 -\frac{M}{r}\right)^2 dt^2 + 
\left( 1 -\frac{M}{r}\right)^{-2}dr^2 + r^2d\theta^2 + r^2\sin^2(\theta)\, d\varphi^2 \,\,\,,\\
T^{a}_{\,\,\,b} &=&  - \frac{M^2}{8\pi r^4}\,{\rm diag}[1,1,-1,-1] \,\,\,,\\
\label{phi3}
\phi &=& \pm\sqrt{\frac{3}{4\pi}} \frac{M}{r-M} \,\,\,.
\end{eqnarray}
As mentioned before, this example is ``in principle''  
covered by T1. However, during the derivation, 
regularity at the horizon was not imposed. The invariants of Sec. III show that 
the only singularity is located at $r=0$. However, the scalar 
field Eq. (\ref{phi3}) has a singularity at the black hole horizon $r=M$, and therefore, 
since $\phi$ is a scalar, it implies that the horizon is not regular. From 
Eq. (\ref{phi0}) or Eq. (\ref{phip}) [cf. also Eqs. (\ref{Treg}) and (\ref{equil0})], 
regularity at the horizon (with $\phi$ and $\partial_r\phi$ bounded there) implies $d=0=a$, 
and the only possibility 
for $\phi$ to be regular is $M=0$, at which case the solution 
reduces to the Minkowski spacetime. This solution has been the issue of 
several controversies that will not be discussed and repeated in detail here. 
Bekenstein \cite{Bek2} has argued that since the tidal effects on test particles 
are bounded at the horizon, the singularity of the scalar field has no physical consequences. 
On the other hand in Refs. \cite{Zannias,sudzan}, it has been argued that the 
pathology of $\phi$ at the horizon reflects the fact that the 
solution is not a genuine solution to the Einstein field equations since it is not 
satisfied at the horizon $r=M$. This has been shown using different regularization 
techniques that lead to different results at the horizon 
\cite{sudzan}. Whether the BBBM solution is or is not a genuine solution of the 
Einstein field equations, what it is clear is that the pathology at the horizon 
is present.

It is to note, that the solutions characterized by T1 are 
genuinely regular as is shown by the invariants, 
except perhaps at the origin, but such a physical 
singularity is to be covered by a regular horizon. The 
theorem T1 states nothing about the matter fields themselves but stems from the
form of the energy-momentum tensor. The ill behavior arises in the BBBM solution 
when integrating the equation for the scalar field and 
the regularity condition is not imposed during the process. This example clearly shows that 
special care has to be taken when the conditions on the energy-momentum 
tensor required by T1 are a priori not satisfied but imposed as to obtain 
the desired form of the solution. 
\bigskip

The following and second example treats another 
solution similar to BBBM but with a regular horizon. 
The example, corresponds to the very recent solution found in 
Ref. \cite{zanelli} using an action like (\ref{jordan}) with $F(\phi)$ 
given also by Eq. (\ref{F}) and taking again $\xi=-1/12$, but with 
$V(\phi)= \lambda \phi^4$ and $\Lambda >0$. The case with $\Lambda=0$
 is also conformal invariant and for such a 
theory no-hair theorems exist \cite{eloy}.

As will be shown in the following, 
the static and spherically symmetric solution of the corresponding field equations 
is the extreme RNdS solution Eqs. (\ref{TSdSRN}) and (\ref{mSdSRN}) 
with $M=Q$ and $C=-Q^2/(8\pi)$, $\lambda= -2\pi\Lambda/9$ 
and with $\phi$ given also by Eq. (\ref{phi3}) \cite{zanelli}. Therefore, the 
solution is covered by theorem T2. To show this, the starting point is the 
Klein-Gordon Eq. (\ref{KGgen}), which reads
\begin{equation}\label{KGzan}
\Box \phi= 4\phi\left(\lambda\phi^2 + \frac{\Lambda}{6}\right)\,\,\,.
\end{equation}
On the other hand, from Eqs. (\ref{effTmunu})$-$(\ref{Geff}), 
and using Eq. (\ref{KGzan}) one has 
\begin{equation}
\label{TECzan}
T_{ab}= \frac{1}{6F}\left\{
4(\nabla_a\phi)(\nabla_b\phi) - 2\phi \nabla_a\nabla_b\phi - 
g_{ab}\left[(\nabla^c\phi) (\nabla_c\phi) - 2V(\phi) + \frac{8\pi}{3}\phi^2 T_{\Lambda}\right] + 6 T_{\Lambda\,\,ab} \right\} \,\,\,.
\end{equation}

One can show that the trace of the above tensor for the scalar field 
alone (i.e., with $T_{\Lambda\,\,ab}=0$), vanishes like in the first example with $V=0$, reflecting the conformal invariance 
of the theory when $\Lambda=0$. One can construct the solution in the same 
way as in the BBBM example by imposing the condition 
$T^{r}_{\,\,\,r}= T^{t}_{\,\,\,t}$. Indeed, from Eq. (\ref{TECzan}) 
the condition $T^{r}_{\,\,\,r}= T^{t}_{\,\,\,t}$ leads to
\begin{equation}\label{phi1zan}
\frac{N(\partial_r\phi)}{r}\partial_r\left(rN\phi\right)= 2V +\frac{\Lambda \phi^2}{3}
\,\,\,\,.
\end{equation}
On the other hand, the Klein-Gordon Eq. (\ref{KGzan}) with $A=N^{-1}$ 
(since $T^{r}_{\,\,\,r}= T^{t}_{\,\,\,t}$) reads
\begin{equation}\label{KGzan2}
\partial^2_{rr}\phi + \frac{2(\partial_r\phi)}{rN}\partial_r\left(rN\right)
= \frac{4\phi}{N^2}\left(\lambda\phi^2 + \frac{\Lambda}{6}\right)\,\,\,.
\end{equation}
From Eqs. (\ref{phi1zan}) and (\ref{KGzan2}) one obtains
\begin{equation}\label{phipp}
\phi\partial^2_{rr}\phi - 2(\partial_r\phi)^2=0\,\,\,.
\end{equation}
This equation is equivalent to Eq. (\ref{phip2}), 
as one can see by differentiating Eq. (\ref{phip2}). Therefore, one solution 
of Eq. (\ref{phipp}) is also given by,
\begin{equation}\label{phi2zan}
\phi= -\frac{e}{r-M}\,\,\,.
\end{equation}
The second solution $\phi={\rm const.}$ will be discussed later.

A straightforward calculation, with the help of Eq. (\ref{phi1zan}), 
shows that the condition Eq. (\ref{newcond}) is verified. 
In addition, Eq. (\ref{phi1zan}) together with Eq. (\ref{phi2zan}), provide 
\begin{equation}\label{NN}
\frac{N^2M}{r} -\left(r-M\right)N\partial_rN= 2\lambda e^2 + \frac{\Lambda}{3}
\left(r-M\right)^2\,\,\,.
\end{equation}
One can verify that
\begin{equation}\label{solN}
N^2= \left(1-\frac{M}{r}\right)^2 - \frac{\Lambda r^2}{3}\,\,\,,
\end{equation}
is the solution of Eq. (\ref{NN}) if and only if
\begin{equation}\label{alpha}
\lambda= -\frac{\Lambda M^2}{6e^2}\,\,\,.
\end{equation}
Finally, from Eq. (\ref{TECzan}) and using Eqs. (\ref{phi1zan}), (\ref{phi2zan}) and 
(\ref{solN}), one obtains
\begin{equation}
T^{t}_{\,\,\,t}= -\frac{e^2 \left(1-\frac{M}{r}\right)
\left(1-\frac{2M}{r}\right)}{6\left[ r^4 \left(1-\frac{M}{r}\right)
\left(1-\frac{2M}{r}\right) + r^2\left(1-\frac{M}{r}\right)\left(M^2 
-\frac{4\pi e^2}{3}\right)\right]} -\frac{\Lambda}{8\pi} \,\,\,.
\end{equation}
Clearly, the choice $e^2=\frac{3 M^2}{4\pi}$ leads to
\begin{equation}\label{Tttzan}
T^{t}_{\,\,\,t}= -\frac{M^2 }{8\pi r^4} -\frac{\Lambda}{8\pi} = T^{r}_{\,\,\,r}\,\,\,.
\end{equation}
and from Eq. (\ref{alpha}), one finds $\lambda= -2\pi\Lambda/9$. This is 
precisely the solution Eq. (\ref{TSdSRN}) with $C=-M^2/(8\pi)$. In 
addition, it is not difficult to show that the T-tensor Eq. (\ref{TECzan}) 
then verifies the condition [cf. Eq. (\ref{newcond})]
$T^{\theta}_{\,\,\,\theta}= - T^{r}_{\,\,\,r} - \frac{\Lambda}{4\pi}\,$ 
needed to obtain the expression Eq. (\ref{Tttzan}) directly from the 
equilibrium Eq. (\ref{equil2}). This shows the full consistency in what 
regards the matter fields. It remains to obtain the solution for $m(r)$.

From Eq. (\ref{solN}) and since $N^2=A^{-2}=(1-2m(r)/r)$ (given that 
$T^{r}_{\,\,\,r}= T^{t}_{\,\,\,t}$) one concludes
\begin{equation}
m(r)= M - \frac{M^2}{2r} + \frac{\Lambda r^3}{6}\,\,\,,
\end{equation}
which is consistent with the expression one would obtain independently 
using Eqs. (\ref{Eeq1b}) and (\ref{Tttzan}). 

Summarizing, as announced, the static and spherically symmetric solution of the 
theory given by Eq. (\ref{jordan}) with a conformal nonminimal coupling scalar field and 
with a positive cosmological constant is given by the extreme 
Reissner-Nordstrom-de-Sitter Eqs. (\ref{mass0ext}) and (\ref{Tabext}) taking 
$\kappa=-1$, $Q=M$, $C=-M^2/(8\pi)$:
\begin{eqnarray}
\label{solzan1}
ds^2 &=& -\left[\left(1-\frac{M}{r}\right)^2 - \frac{\Lambda r^2}{3}\right]dt^2 
+ \left[\left(1-\frac{M}{r}\right)^2 - \frac{\Lambda r^2}{3}\right]^{-1} 
dr^2 + r^2d\theta^2 
+ r^2\sin^2(\theta)\, d\varphi^2 \,\,\,\,,\\
\label{Tttzan2}
T^{a}_{\,\,\,b} &=& -\frac{M^2}{8\pi r^4}\,{\rm diag}[1,1,-1,-1]
- \frac{\Lambda}{8\pi} \,{\rm diag}[1,1,1,1] \,\,\,,\\
\label{solzan2}
\phi &=& \pm \sqrt{\frac{3}{4\pi}}\frac{M}{r-M}\,\,\,.
\end{eqnarray}
and with $V(\phi)= -\frac{2\pi\Lambda}{9}\phi^4$.

Unlike the BBBM solution, this solution is regular in the domain of outer communication, 
since none 
of the horizons coincide with $M$. Moreover, the singularity of the scalar field 
at $M$ is hidden 
by the event horizon located at $r_h= \frac{l}{2}\left(1-\sqrt{1-4M/l}\right)$ 
where $l=\sqrt{3/\Lambda}$. The static BH 
solution is then restricted to $0< M <l/4$, $\Lambda>0$ \cite{zanelli}. Note 
that in the limit $\Lambda\rightarrow 0$, it turns out $V(\phi)\rightarrow 0$ and the BBBM solution 
is recovered.

Another solution of Eqs. (\ref{phi1zan}) and (\ref{KGzan2}) is \cite{zanelli}
\begin{equation}\label{phi3zan}
\phi^2= -\frac{\Lambda}{6\lambda}= {\rm const.} \,\,\,
\end{equation}
provided $\Lambda/\lambda <0$. Using Eq. (\ref{phi3zan}) in Eqs. (\ref{F}) and 
(\ref{TECzan}) one finds
\begin{eqnarray}
T_{ab} &=& -\frac{\Lambda}{8\pi}\,\,g_{ab} \,\,\,,\\
G_{eff} &=& \frac{1}{1 + \frac{2\pi \Lambda}{9\lambda}}\,\,\,,
\end{eqnarray}
where Eq. (\ref{Geff}) was also used. A positive $G_{eff}$ (together with 
the condition $\Lambda/\lambda <0$) demands therefore 
$-\frac{9}{2\pi} < \frac{\Lambda}{\lambda} <0 $.
The static and spherically symmetric solution to Einstein's field equations is 
then given by the \sch$\!\!$-dS/AdS solution.

This example like the previous one shows that the theorem T1 or its trivial 
extension T2 are 
useful to understanding in a rather systematic and straightforward fashion 
the existence of solutions like the ones analyzed above.
\bigskip

Finally, the theorem T1 helps also to clarify certain confusions \cite{Edery}, about the 
form of the canonical metric Eq. (\ref{metric}) used to describe 
some kind of static and spherically symmetric spacetimes (see also the comment of Ref. \cite{Bek3}). 
Namely, the spacetime generated by the dark matter in spiral galaxies.
 According to T1, one can assume $N=A^{-1}$, if and only if 
$T^{t}_{\,\,\,t}= T^{r}_{\,\,\,r}$. As is showed in Refs. 
\cite{us,lesgourgues}, 
one can, in principle, model the dark matter in galactic halos with an 
(inhomogeneous) perfect fluid 
or scalar fields. For those matter fields $T^{t}_{\,\,\,t}\neq T^{r}_{\,\,\,r}$, and 
therefore $N\neq A^{-1}$, at least in the region where the dark halo is presumably 
the responsible for the so-called flat rotation curves. That analysis does not contradict at 
all the fact that in the vicinity of a star (like the Sun) within the galaxy, 
where the curvature produced by the dark halo is 
very small as compared with the curvature generated by the star, the metric 
is to a very good approximation given by the \sch metric $N_{\odot}= 
A^{-1}_\odot= 1-2M_\odot/r$, which successfully reproduces the solar system experiments. 
In other words, the laws of gravity are the same 
(given by the Einstein field equations) but the same laws can produce distinct 
effects due to the concentration of different kinds of matter at different scales.

\section{Conclusion}
A simple theorem based on certain restrictions on the energy-momentum tensor was 
proved, showing a large variety of two-parameter static and spherically symmetric 
black hole solutions. Among them, the most known exact black hole solutions and 
some recent ones are included. A trivial extension of the theorem allows one to cover 
a three-parameter family of solutions, with one of the parameters 
being a cosmological constant. An issue that remains to 
be investigated is if there exist indeed physical fields 
or some field theory behind some of the new solutions proposed by the theorem, 
as the latter does not include 
the specific form of those fields. In the same way, one can also ask 
(in the spirit of the no-hair theorems or no-hair conjectures) if two different energy-momentum tensors (that satisfy however the conditions of the theorem) 
can generate the same 
spacetime (same $\kappa$). That is, if the corresponding spacetimes can be identified 
with each other (belonging to the same family) but with the constant $C$ playing different 
roles in each theory.  If 
it were not by the fact that the scalar field is not well defined at the 
horizon, the BBBM solution Eqs. (\ref{solBBBM})$-$(\ref{phi3})
which is contained within the 
Reissner-Nordstrom family but with a scalar field instead of the electric field, 
would undisputedly answer the question in the affirmative. On the other hand, 
the solution Eqs. (\ref{solzan1})$-$(\ref{solzan2}) \cite{zanelli}, it does represent the same spacetime that the 
extreme RNdS solution but with a different energy-momentum tensor and moreover, it 
does not show the same pathology as the BBBM solution. Therefore, this solution 
seems to truly answer the question positively \cite{refs}.

\bigskip
{\bf Acknowledgments}
\bigskip

I thank U. Nucamendi for helpful comments and J.A. Gonz\'alez for some assistance with 
Maple. This work was partially supported by Conacyt grant {\tt 32551-E} and 
DGAPA-UNAM grants {\tt IN112401} and {\tt IN122002}.


\begin{thebibliography}{}
\bibitem{comment0}
Incidentally, by allowing the value $\kappa=1$ in Eqs. (\ref{mass0ext}) and 
(\ref{Tabext}), 
one could have ``two'' cosmological 
constants $\Lambda$ and $C:= -\Lambda_c/(8\pi)$ which is equivalent to 
an effective cosmological constant $\Lambda_{eff}= \Lambda + \Lambda_c$.
\bibitem{comment}
If some concrete examples of energy-momentum tensors were found such that the 
spacetimes generated by them are not asymptotically flat 
(i.e., for $\kappa \geq -1/2$), 
one should verify (apart from the 
cases $\kappa=0,1$ where black holes can be defined) that 
the notion of $\mathscr{I}^{+}$ exists for those spacetimes in order for black holes 
to be well understood.
\bibitem{commentmon}
For $\kappa=0$, the metric Eq. (\ref{gab}) writes  
$ds^2= - \left[ 1 - \Delta - 
\frac{2M}{r}\right] dt^2 + \left[ 1 - \Delta- \frac{2M}{r}\right]^{-1} 
dr^2 +  r^2d\theta^2 + r^2\sin^2(\theta)\, d\varphi^2 $. By performing the 
coordinate transformation $r\rightarrow (1-\Delta)^{1/2}\,r$ and 
$t\rightarrow (1-\Delta)^{-1/2}\,t$, together with $M=M_{\rm ADM_\Delta}(1-\Delta)^{3/2}$ 
the metric reads $ds^2= - \left[ 1 - 
\frac{2M_{\rm ADM_\Delta}}{r}\right] dt^2 + \left[ 1 - \frac{2M_{ADM_\Delta}}{r}\right]^{-1} 
dr^2 + (1-\Delta)r^2 \left(d\theta^2 + \sin^2(\theta)\, d\varphi^2\right) $, which shows explicitly that the spacetime is asymptotically flat 
with a solid angle deficit.
\bibitem{comment2}
The solution of Bekenstein (see Ref. \cite{Bek1}) was found originally using a conformal 
method (and with an electric charge) rather than following the approach presented here.
\end{thebibliography}
\end{document}